\documentstyle[apalike,amssymb,epsf,12pt]{article}
\def\a{\alpha}
\def\b{\beta}

\def\g{\gamma}
\def\d{\delta}
\def\e{\varepsilon}

\def\t{\theta}
\def\C{{\bf C}}
\def\D{\Delta}
\def\S{{\bf S}}

\begin{document}


\title{{\bf Statistical thermodynamics of membrane bending mediated
protein-protein attractions}}

\author{Tom Chou$^{1}$\footnote{Correspondence: 
Mathematics Dept. 382P, Stanford University, Stanford, CA 94305, 
email: tc@math.stanford.edu}, Ken S. Kim$^{2}$, and 
George Oster$^{3}$ \\
$^{1}$Mathematics Dept. Stanford University, \\
$^{2}$Dept. of Physiology and DAMTP, University of Cambridge \\
$^{3}$Dept. of Molecular and Cell Biology, University of California, Berkeley}

\date{\today}
\maketitle

\begin{abstract}

Integral membrane proteins deform the surrounding bilayer
creating long-ranged forces that influence distant proteins. 
These forces can be attractive or repulsive, depending on the
proteins' shape, height, contact angle with the bilayer, as
well as the local membrane curvature.  Although interaction
energies are not pairwise additive, for sufficiently low
protein density, thermodynamic properties depend only upon pair
interactions.  Here, we compute pair interaction potentials and
entropic contributions to the two-dimensional osmotic pressure
of a collection of noncircular proteins.  In contrast to direct
short-ranged interactions such as van der Waal's, hydrophobic,
or electrostatic interactions, both local membrane Gaussian
curvature and protein ellipticity can induce attractions
between two proteins at distances of up to ten times their
typical radii. For flat membranes, bending rigidities of $\sim
30k_{B}T$, and moderate ellipticities, we find  thermally
averaged attractive interactions of order $\sim 2k_{B}T$. 
These interactions may play an important role in the
intermediate stages of protein aggregation.

\end{abstract}



\vspace{2mm}


\section{Introduction}

Membrane proteins interact directly via screened
electrostatic, van der Waal's, and hydrophobic forces. 
These are short ranged, operating typically over distances
of less than a nanometer.  Proteins can also interact
indirectly via the bilayer in which they are dissolved. In
particular, a protein that is ``geometrically mismatched''
to the bilayer will induce deformations that affect
neighboring proteins.  These ``solvent induced forces'' (the
membrane lipids being the solvent) are generated by
bending deformations of the  bilayer and typically act
over many protein diameters.

Membrane associated proteins can aggregate due to bilayer
bending mediated interactions. For example, aquaporin AQP1 and
CD59 aggregate to tips of pipette-drawn tubules
\cite{CHO,DISCHER}. Previous studies using a continuum
approximation for the intervening bilayer membrane, have
treated protein-protein interactions and found an $r^{-4}$
repulsion between two identical inclusions
\cite{GOULIAN2,KIM,PARK,FOURNIER}.  Goulian {\it et
al.}\cite{GOULIAN2} also find a weak attractive
($-k_{B}T/r^{4}$) interaction arising from Casimir forces
resulting from suppressed thermodynamic fluctuations of the
intervening membrane. Here, we study in detail a direct
mechanical origin for protein-protein attractive interactions. 
Although bending induced forces between multiple inclusions are
not pairwise additive, \cite{KIM,KIM2}, in this paper, we
restrict ourselves to low protein densities where only pairwise
interactions are relevant.  We find that the interplay between
protein noncircularity \cite{KIM2} and background Gaussian
curvature curvature dramatically affect protein-protein
attractions and thermodynamics.

Many membrane proteins are noncircular in the plane of the
membrane, including adsorbed polypeptides such as MARCKS
\cite{MARCKS}, and bacteriorhodopsin \cite{RHODOPSIN}, which
consists of seven transmembrane helices arranged in an
elliptical configuration.  Small domains, dimers, or
droplets of molecules such as cholesterol or specific lipids
can themselves behave effectively as membrane inclusions. 
Droplets need not be rigid to induce attractive forces among
themselves; fluctuations in the droplet shape itself may
lead to an effective attraction.

In Section 2 we briefly review the mechanical theory of inclusion-induced
bilayer bending \cite{HELFRICH,KIM,NETZ}. The lipid membrane is
approximated by a thin plate that resists out-of-plane bending. 
Inclusions such as integral membrane proteins, or surface adsorbed
molecules, impose boundary conditions along the contact line between the
membrane and the protein. Using elastic plate theory to describe the
membrane deformations, we derive the energy for two identical inclusions as
a function of their relative position within the membrane surface.

In Section 3, we show that the rotational and translational time
scales can be separated, so that we can thermally average out the
fast rotational degrees of freedom. The resulting effective potential
between two proteins is attractive provided the inclusions are
sufficiently non-circular. We use the effective potential to compute
the second virial coefficient and  show how the attractive
interactions affect the two-dimensional protein osmotic pressure. 
Finally, we discuss biological processes where membrane induced
long-ranged protein-protein attractions may play an intermediate
role, and propose possible measurements.

\section{Membrane inclusions and height deformation}

Small membrane deformations (on the scale of the lipid or protein
molecules) can be accurately modeled using standard plate theory
\cite{ELASTICITY,HELFRICH}

\begin{equation}
\tilde{E}[H(\S),K(\S)] = 2b\oint d\S H^{2}(\S) + b_{g}\oint d\S K(\S),
\label{G}
\end{equation}

\noindent where $H(\S)$ and $K(\S)$ are the local mean and
Gaussian
curvatures, and $b$ and $b_{g}$ are their associated elastic moduli.  We
have assumed a symmetric bilayer and a vanishing spontaneous mean
curvature in the absence of the membrane-deforming proteins of interest. 
For uniform $b_{g}$, the Gaussian contribution (the second integral in
Eq.  \ref{G}), when integrated over the entire surface yields a constant
that is independent of the relative configurations of the embedded proteins
\cite{KIM,STRUIK}.  Thus, the Gaussian energy term can be ignored when
considering protein-protein interaction energies.

In an expansion of the free energy about that of a flat
interface, $H(\S) \simeq (1/2)\nabla^{2}h(x,y)$, where
$\nabla^{2}$ is the two-dimensional, in-plane Laplacian, and 
$h(x,y)$ is a small, slowly varying height deformation from the
flat state (cf. Fig 1).  Minimizing $\tilde{E}[h(\S)]$ with
respect to $h(x,y \in \S)$ yields the biharmonic equation

\begin{equation}
\nabla^{4}h(x,y) = 2\nabla^{2}H({\bf r}) = 0.
\label{BIHARMONIC}
\end{equation}

\begin{figure}
\begin{center}
\leavevmode
\epsfxsize=4.8in
\epsfbox{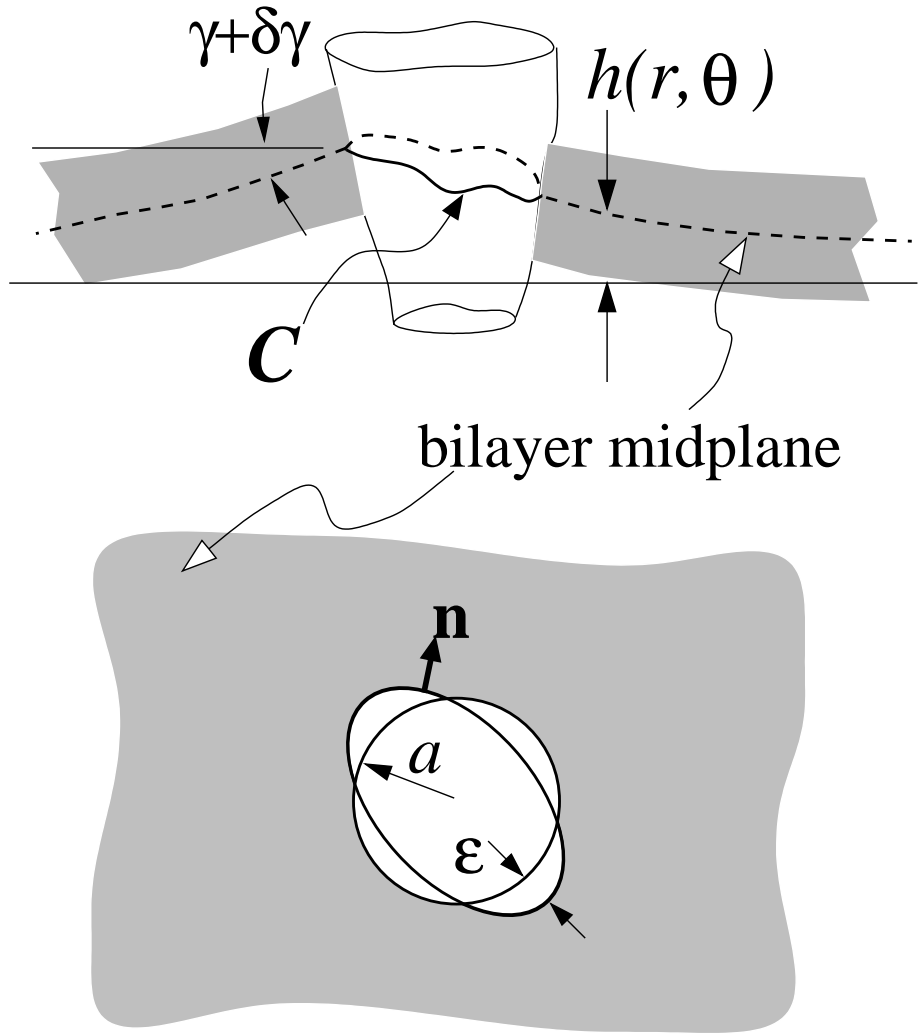}
\end{center}
\caption{Schematic of a protein inclusion. The top 
figure is a cut-away view of a membrane protein 
that contacts the continuum bilayer midplane on curve ${\bf C}$.
The contact angle on ${\bf C}$ is denoted $\gamma + \delta\gamma$, while the 
bilayer deviation from a reference flat state is $h({\bf r})$.
The bottom picture shows a possible ellipticity $\varepsilon$
in the projection of ${\bf C}$ onto the midplane.}
\label{FIG1}
\end{figure}

\noindent First consider membrane deformations about an isolated,
circularly symmetric inclusion of radius $a$. If the bilayer midplane
contacts\footnote{The contact angle $\gamma$ incorporates the details of
the molecular interactions between the included/adsorbed protein with the
lipid molecules. A molecular dynamics simulation would in principle
provide the basis for a quantitative estimate of $\gamma$, but is beyond
the scope of this paper. We will simply assume that $\gamma$ is a
phenomenological parameter determined by the local chemistry, in complete
analogy with the standard liquid-gas-solid contact angle.}  the protein
perimeter ${\bf C}$ (see Fig. \ref{FIG1}) at an angle $\gamma$, the
appropriate solution to Eq.  \ref{BIHARMONIC} is $h(r) = -\gamma\ln
(r/a)$ for $r>a$.  We have excluded terms in $h(r)$ of the form $r^{2}\ln
r, r^{2}, \mbox{const.}$ because they are unbounded in energy (Eq. 
\ref{G}), or do not satisfy the contact angle boundary condition at
$r=a$.  Since $\nabla^{2} \ln (r/a) = 2H(r) = 0$ for $r>a$, there is no
mean curvature bending energy (proportional to $b$) residing in the
bilayer.

Now consider cases where more than one inclusion are present,
or where the contact angles, heights of contact, or the shapes
of the membrane associated proteins are noncircular.  Three
types of noncircularity can arise. The inclusion itself may be
noncircular ({\it e.g.} elliptical), the height of contact of
the bilayer midplane to the inclusion may vary along the
perimeter $\C$ of the protein, and the contact angle itself may
vary along $\C$. These noncircular boundary effects arise from
the detailed microscopic nature of the protein and its
interaction with the lipid molecules. When more than one
protein is present, the deformations surrounding each protein
are not circularly symmetric.  A nonvanishing mean curvature,
$H({\bf r})$, that gives bounded bending energies can be
represented by a multipole expansion,

\begin{equation}
H(r,\theta) = \sum_{n=2}^{\infty}r^{-n}\left(
a_{n}\cos n\theta + b_{n}\sin n\theta\right),
\label{KAPPAMP}
\end{equation}

\noindent where $(r, \theta)$ is the radial and angular
coordinate about an arbitrary origin.  Upon substitution of Eq.
\ref{KAPPAMP} into Eq.  \ref{G}, we find the bending energy
\nobreak{$\tilde{E} \sim
b\sum_{n=2}^{\infty}(a_{n}^{2}+b_{n}^{2})$.}  To determine
$a_{n}, b_{n}$, we solve $\nabla_{\perp}^{2}h(r,\theta) =
H(r,\theta)$ and impose boundary conditions (see Appendix A) on
$h(r,\theta)$ at ${\bf C}$.  In the limit of small
noncircularity or low protein concentrations, the largest
nondivergent terms are associated with $n=2$.  Wiggly inclusion
cross-sections or highly oscillating boundary conditions only
weakly affect membrane bending-mediated protein-protein
interactions via $n>2$ terms.  We derive the full multibody,
interaction energy in Appendix A.  The {\it two-body}
interaction energy measured in units of $k_{B}T$ is

\begin{equation}
E(R,\theta_{1},\theta_{2};\D, K_{b},\Omega) = \Big| {e^{-2i\Omega}\over
R^{2}}+K_{b} 
- {\D\over 2}e^{-2i\theta_{1}}\Big|^{2} + \Big| {e^{-2i\Omega}
\over R^{2}}+K_{b} - {\D\over
2}e^{-2i\theta_{2}}\Big|^{2}. 
\label{U2}
\end{equation} 

\noindent The dimensionless separation distance $R$,
protein ellipticity $\D$, and background curvature $K_{b}$ are given by

\begin{equation}
\begin{array}{l}
\displaystyle R \equiv {r\over R_{0}}, \quad R_{0} \equiv a\sqrt{\gamma}B^{1/4} \\[13pt]
\displaystyle \D \equiv \bar{\e}\sqrt{B}
\\[13pt]
\displaystyle K_{b} \equiv a\sqrt{B}\left({\partial^{2} 
h_{b}(x_{1},x_{2})\over \partial x_{1}^{2}}\right),
\end{array}
\label{RESCALING}
\end{equation}

\noindent where $B \equiv \pi b/k_{B}T$ is the dimensionless
bending stiffness, and $\bar{\e} \sim O(\e)$ quantifies the
noncircular nature of each inclusion (see Appendix A). The
angle $\Omega$ is measured between the line joining the protein
centers and the principle axis of curvature defined by the
background Gaussian curvature (see Fig.  \ref{COORD}).  The
angles $\theta_{1},\theta_{2}$ are measured between the
principle axes of proteins 1 and 2 and the same principle axis.
The quantity $K_{b}$ measures the local, externally induced
(via other distant proteins or external bending forces)
background curvature in this principle axis direction.  We show
in Appendix A that the dominant effect of distant
proteins is to induce mean curvature deformations that decay as
$1/r^{2}$, but constant negative Gaussian curvatures. The local
curvature $K_{b}$ arises only from deformations that are of
zero mean curvature. In what follows, our statistical
thermodynamic analyses will be applied to the pair interaction
energy given by Eq.  \ref{U2} with the convention $\D, K_{b}
\geq 0$.  

\section{Rotationally averaged interactions}

Proteins that  are not attached to the cytoskeleton are free to
rotate and diffuse within in the membrane.  The interaction
potential between two membrane-deforming inclusions is a
complicated, nonseparable function of their relative major axis
angles and separation distance (cf. Eq.  \ref{U2}). Although
the energy is a function of the specific separations and angles
between two membrane associated proteins, rotational degrees of
freedom are sampled faster than the translational of freedom. 
This can be shown by the following argument.

A small solvent molecule in solution has a rotational
correlation time of the order $\tau_{\em rot}\lesssim 1$ ns,
while its translational diffusion constant is $D_{\em trans}
\sim 10^{-6}$cm$^{2}$/s.  Therefore, in the time $\tau_{\em
rot}$ it takes for a small solvent molecule to lose rotational
correlation, it would have translated

\begin{equation}
\d r \sim \sqrt{\tau_{\em rot}D_{\em trans}} \sim 0.1\mbox{nm}.
\end{equation}


\noindent For membrane constituents, such as bilayer lipid
molecules, $\tau_{\em rot}\sim 1-5$ns, and $D_{\em trans} \sim
10^{-7}$cm$^{2}$/s, where $\tau_{\em rot}$ corresponds to
rotation about the molecular axis parallel to the normal vector
of the membrane \cite{MARSH}.  As with small molecules in bulk
solution, membrane-bound lipid molecules also move $\d R \sim
0.1\mbox{nm}$ during a rotational correlation time.  For larger
membrane inclusions such as proteins, both rotation and
translational diffusion are slower. If $\Lambda$ is the
relative size of the protein radius with respect to the
effective lipid radius, protein rotational correlation times
increase by $\sim \Lambda^{3}$ while $D_{\em trans}$ decreases
with $a$.  Membrane proteins that are not too large,
$\Lambda\lesssim 10$ say, diffuse $\d r \sim 1$nm during the
time over which it has lost rotational correlation.  Therefore,
in the time it takes for a typical inclusion to rotate about
its axis, it has diffused less than its own size,
typically $\gtrsim$1nm.  This estimate is consistent with
fluorescence measurements that find $\tau_{\em rot}\sim
0.1-1$ms \cite{TAUROT}. Rotational time scales for larger
proteins may not be much faster than translational motions,
therefore, our subsequent model is most appropriate for small,
unhindered membrane proteins. We must eventually verify that
the protein-protein separation $r$ of interest is
greater than the typical diffusion distance $\d r$.

The time scale separation can be implemented by statistically
averaging over the principle axis angles of the two inclusions
while keeping the distance $R$ and angle $\Omega$ between them
fixed. Weighting the exact two particle energy over its own
Boltzmann weight,

\begin{equation}
E_{\em eff}(R;\D, K_{b},\Omega) = Z^{-1}\int_{0}^{2\pi}
E(R;\D,K_{b}\theta_{1},\theta_{2})e^{-
E(R,\theta_{1},\theta_{2};\D, K_{b},\Omega)}d\theta_{1}d\theta_{2},
\label{EEFF}
\end{equation}

\noindent where the rotational partition function 

\begin{equation}
Z \equiv 
\int_{0}^{2\pi} e^{-
E(R,\theta_{1},\theta_{2};\D, K_{b},\Omega)}d\theta_{1}d\theta_{2}.
\label{Z}
\end{equation}

\noindent Upon substitution of Eq. \ref{U2} into Eqs.
\ref{EEFF} and \ref{Z}, and performing the integration (see Appendix B), 

\begin{equation}
E_{\em eff}(R;K_{b},\Omega) = {2\xi^{2}\over \D^{2}} +{\D^{2}\over 2}- 2\xi{I_{1}
(\xi)\over I_{0}(\xi)}
\label{UEFF2}
\end{equation}

\noindent where

\begin{equation}
\begin{array}{l}
\displaystyle \xi  = \D\sqrt{{1\over R^{4}}+ {2K_{b}\over
R^{2}}\cos 2\Omega + K_{b}^{2}}
\label{XI}
\end{array}
\end{equation}

\noindent The effective interaction potential 
between two inclusions is defined by the difference 
between the membrane bending energies 
of two inclusions separated at distance $R$ and 
at infinite separation,

\begin{equation}
\begin{array}{rl}
\displaystyle U_{\em eff}(R;\D,K_{b},\Omega) & = E_{\em
eff}(R;\D,K_{b},\Omega)-E_{\em eff}(\infty)\\[13pt]
\: & \displaystyle \equiv
{2\xi^{2}\over \D^{2}} - {2\xi I_{1}(\xi)\over I_{0}(\xi)} - \left[2K_{b}^{2}
-2\D K_{b}{I_{1}(\D K_{b})\over I_{0}(\D K_{b})}\right],
\end{array}
\label{UEFF} 
\end{equation}

\noindent For fixed ellipticity $\D$, the set of parameters $K_{b},\Omega$ and $R$ that
gives rise to a minimum at $R^{*} < \infty$,  if it exists, is implicitly
determined by 

\begin{equation}
\left({\partial U_{\em eff}\over 
\partial R}\right)_{R^{*}} =0.
\label{MINROOT}
\end{equation}

\noindent For sufficiently small $R$, $U_{\em eff} \simeq 2/R^{4}$, as in the circular
protein case.

\subsection{Zero background curvature} 

First consider the case of two isolated proteins embedded in a
flat membrane.  In the absence of external mechanical forces
that impose background membrane deformations, and with other
inclusions sufficiently far away, $H_{b}=K_{b} = 0$, and $\xi =
\vert\D\vert/R^{2}$. The effective potential (Eq. \ref{UEFF})
becomes

\begin{equation}
U_{\em eff}(R;\D,K_{b}=0) = {2\over R^{4}}- 
\left({2\D\over R^{2}}\right){I_{1}(\D/R^{2})\over
I_{0}(\D/R^{2})}.
\label{UEFFK=0}
\end{equation}

Without background curvature $(K_{b} = 0)$, there are no
defining principle axes, and $U_{\em eff}$ is independent of
how the angle of the segment joining the inclusion centers is
aligned.  Clearly, an effective attractive interaction can
arise for $\D/R^{2} \gg 1$, when
$I_{1}(\D/R^{2})/I_{0}(\D/R^{2}) \sim 1$, and $U_{\em
eff}(R;\D,K_{b}=0) \sim 1/R^{4} - \D/R^{2}$. Although the
interaction (Eq.  \ref{U2}) yields both repulsive
as well as attractive forces, the Boltzmann thermal average in
Eq.  \ref{EEFF} favors the lower energy configurations of
$\theta_{1},\theta_{2}$.  Hence the pair of inclusions spends
more time in attractive configurations, resulting in a residual
attraction in $U_{\em eff}(R)$. In the $K_{b} = 0$ limit, the
large $R$ behavior of Eq. \ref{UEFFK=0} is

\begin{equation}
U_{\em eff}(R) = {2-\D^{2}\over R^{4}} + O(R^{-6}).
\end{equation}

\noindent Since the potential becomes repulsive at short distances, 
an effective ellipticity $\D > \D^{*} \equiv \sqrt{2}$ is necessary
for the existence of a minimum in $U_{\em eff}(R)$. 
 
Figure \ref{RMIN0/UMIN0}a shows the $\theta-$independent
effective interaction potential as a function of $R$ for
various $\D$.  As $\D$ is increased from $\D^{*}=\sqrt{2}$, the
minimum radius $R^{*}$ determined by Eq.  \ref{MINROOT},
decreases rapidly from $R^{*} \sim \infty$. The $\D > \D^{*}$
dependence of $R^{*}$ is plotted in Figure \ref{RMIN0/UMIN0}b. 
Also shown are the corresponding magnitudes of the global
minima of $U_{\em eff}(R;\D,K_{b}=0)$ as a function of $\D$.  

Figures \ref{RMIN0/UMIN0}a,b show that appreciable attractive
wells can persist at distances $R\sim 1$.  For example, the
minimum determined by the set of parameters $\D\approx 2,\,
R^{*}\approx 0.9$ and $\vert U_{\em eff}(R^{*}) \vert \approx 2
(k_{B}T)$ can arise for $\gamma \sim 1,\, \e/a \sim 0.25$ and
$b \simeq 30k_{B}T$.  The separation corresponding to the
minimum energy in this case is $r^{*}/a \simeq 9$, or nine
times the inclusion radius. Our initial assumptions using
measured and estimated rotational/translational diffusion
constants for typical membrane proteins are validated since $\d
r/a \sim 1-2 \ll r^{*}/a$.  We conclude that thermally averaged
noncircular membrane deformations can induce long-ranged
attractive interactions of at least $\sim 2k_{B}T$ up to
distances  $\sim 10a$.

\subsection{Effect of local Gaussian curvature, $H_{b}=0, K_{b} \neq 0$} 

A local background curvature may arise due to a nonuniform distribution
of distant membrane proteins, or an externally imposed deformation. For
example, in the experiments of \cite{CHO,DISCHER}, a cell is manipulated
by a micropipette. A lipid neck is drawn into the pipette and curvature
is externally imposed. Regions near the base of the neck will have a
large negative Gaussian curvature.  Similarly, membrane fusion and
fission processes in endo/exocytosis involves intermediate shapes with
constricted necks containing Gaussian curvature. These regions may be
``externally'' imposed by proteins involved in vesiculation ({\it e.g.}
dynamin or motor proteins).  The Gaussian curvature in this case may also
result from lipid structural or composition changes \cite{NATURE}.
Therefore, curvature can couple to membrane protein or lipid shapes.

\begin{figure}
\begin{center}
\leavevmode
\epsfxsize=6.2in
\epsfbox{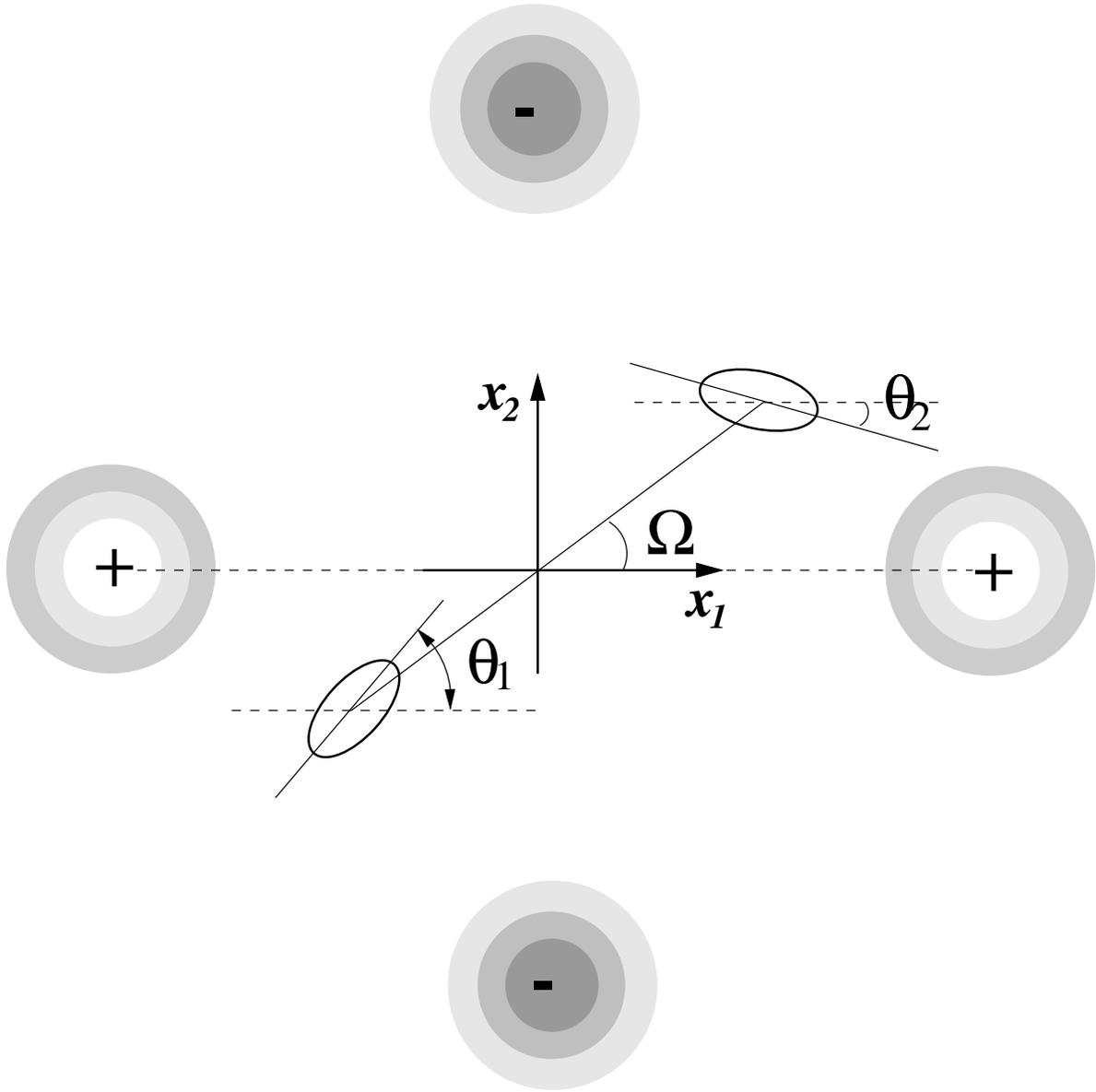}
\end{center}
\caption{Two inclusions embedded in a local saddle deformation. The 
$+/-$ correspond to raised/depressed regions of the membrane. 
The principle axis is aligned with the path joining the two raised regions
(east-west). 
The principle axes of the inclusions $(\theta_{1},\theta_{2})$ as well 
as the centerline joining their centers $(\Omega)$
are measured with respect to this principle axis.}
\label{COORD}
\end{figure}

\begin{figure}
\begin{center}
\leavevmode
\epsfxsize=5.1in
\epsfbox{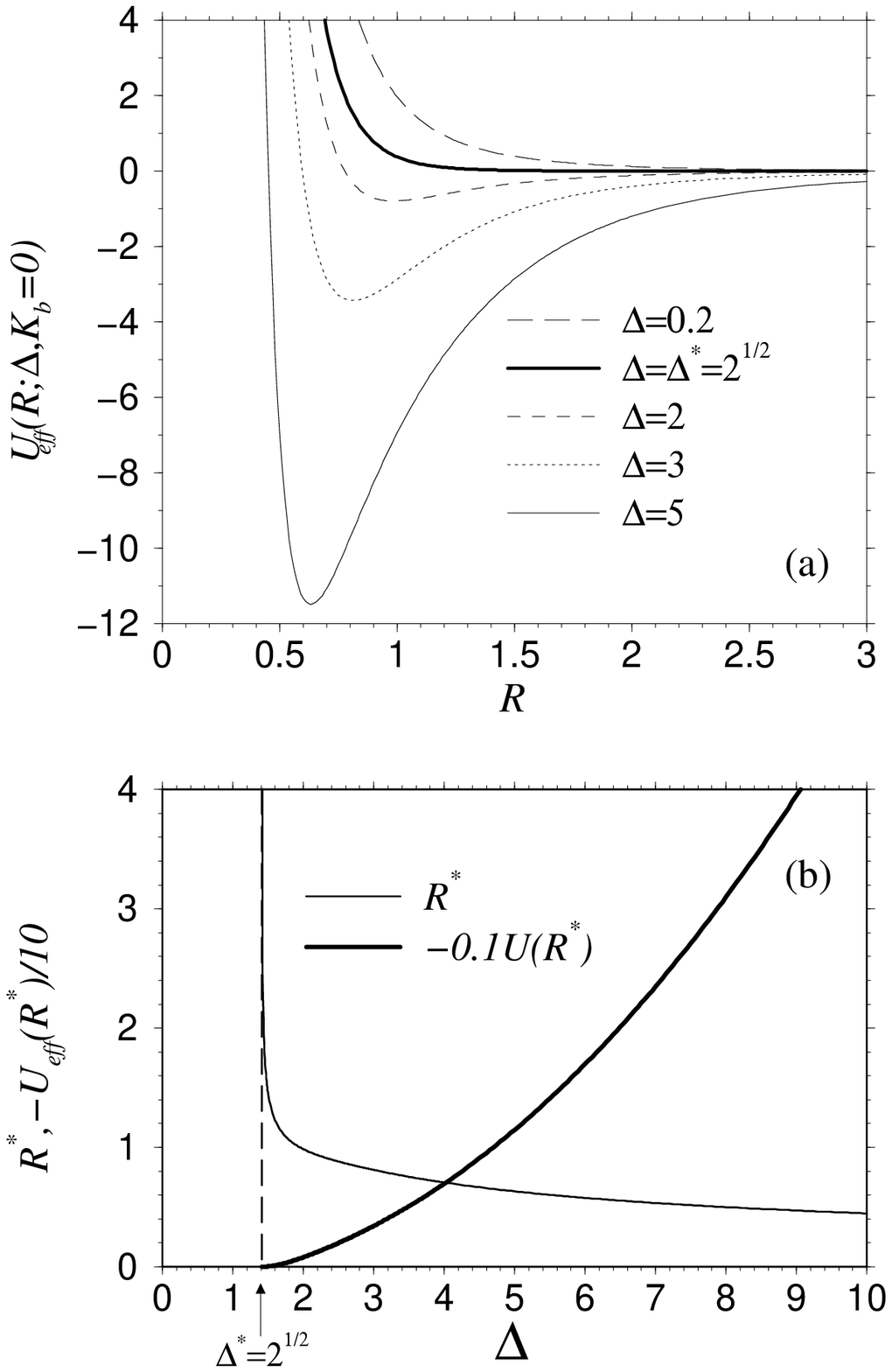}
\end{center}
\caption{(a). Rotationally averaged effective potential (Eq. 
\ref{UEFFK=0}) as a function of protein separation in  a flat membrane
($H_{b}=K_{b}=0$).  (b). The minimum effective energy and its
associated radius $R^{*}$. The minimum of the potential is
plotted as 1/10$\vert U_{\em eff}(R^{*})\vert$. Note that
$R^{*}$ quickly decreases when $\D$ increases above
$\D^{*}=\sqrt{2}$. For large $\D\gg 1$, $R^{*} \sim
\sqrt{2/\D}$ and $\vert U_{\em eff}(R^{*})\vert \sim \D^{2}/2$.}
\label{RMIN0/UMIN0}
\end{figure}

The Gaussian curvature of the membrane between the two proteins
establishes local axes of principle curvature such that
$a\partial_{x_{1}}^{2}h(x_{1},x_{2}) =
-a\partial_{x_{2}}^{2}h(x_{1},x_{2}) \equiv \eta_{b} \propto K_{b} > 0$. 
Since we assume $H_{b} = 0$, the background deformation between the two
proteins will resemble a saddle with principle curvatures of equal
magnitudes (cf. Fig. \ref{COORD}).  The rotationally averaged (over
$\theta_{1},\theta_{2}$) effective interaction $U_{\em eff}(R;\D,
K_{b},\Omega)$ will generate attractions at specific orientation angles
$\Omega$ even if $\D < \D^{*}$. This can be most easily seen by expanding
equation \ref{UEFF} (the rotationally averaged interaction $U_{\em
eff}(R; \D,K_{b},\Omega)$) in powers of $1/R$ for large $R$:

\begin{equation}
U_{\em eff}(R\rightarrow \infty; \D,K_{b},\Omega) \simeq
{A_{2} \over R^{2}}\cos 2\Omega + 
{A_{4} \over R^{4}} + O\left(R^{-6}\right),
\label{EXPANSIONR}
\end{equation}

\noindent where explicit forms for $A_{2}, A_{4}$ are given in Appendix
A. The appearance of $A_{2}\neq 0$ when $K_{b} > 0$ immediately generates
a minimum, however small. Even when ellipticity vanishes ($\D=0$),
$A_{2}\propto  K_{b}\cos 2\Omega < 0$ for appropriate $\Omega$.  

The physical origin of attractions in the presence of background
curvature can be readily seen  by considering Figure \ref{COORD}. With
our convention $\gamma > 0$, circular proteins situated at low regions
of the saddle ($\Omega \sim \pi/2$) develop attractive interactions,
while those with $\Omega \sim 0$ always repel. Recall from previous
studies that two circular protein repel with a $R^{-4}$ potential
\cite{GOULIAN2,KIM,PARK,FOURNIER}.  This is a direct consequence of
placing a second protein in the Gaussian curvature of the first one. 
When the background curvature of the membrane in the region between
two proteins augments the individual Gaussian curvatures around the
first protein (near $\Omega=0$), the $R^{-4}$ repulsion is also
enhanced. Conversely, if the background curvature mitigates the saddle
induced by an individual inclusion (near $\Omega =\pi/2$), the other
inclusion sees not only a diminished repulsion, but a mutual
attraction at large enough distances.  This is because the individual
Gaussian curvature around a protein (arising from $h(r) \approx
-\gamma \ln (r/a)$) decays as $1/r^{4}$ and eventually becomes smaller
than the imposed constant background Gaussian curvature associated
with $K_{b}$.  Attractive effects of the background curvature
eventually manifest themselves when $\Omega \sim \pi/2$.

\begin{figure}
\begin{center}
\leavevmode
\epsfysize=7.4in
\epsfbox{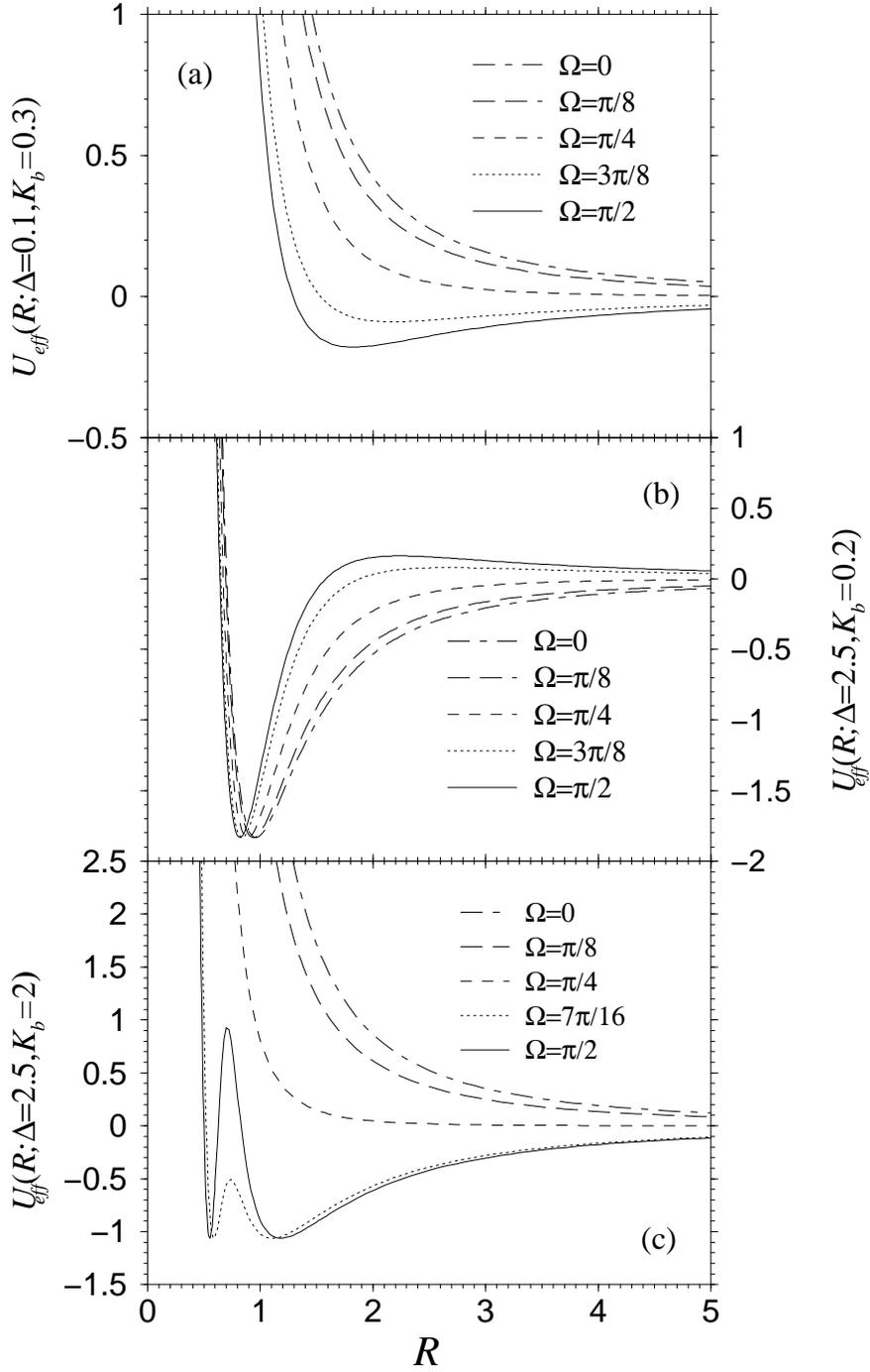}
\end{center}
\caption{Effective potentials between two inclusions embedded in an $H_{b}=0$
and constant $K_{b}$ membrane. (a). $K_{b} = 0.3; \D=0.1$
for various  $\Omega$. (b). $K_{b} = 0.2; \D = 2.5$, and (c). 
$K_{b} = 2; \D = 2.5$. This latter case, although rare under physiological 
conditions, yields two energy minima which are physical manifestations of 
the qualitatively different minima depicted in (a). and (b).}
\label{UOm}
\end{figure}

Figure \ref{UOm}a shows the effects of a small amount of local
background curvature on the effective interaction potential. For
small ellipticity $\D \ll \D^{*}$, minima appear for large enough
angles $\Omega$ (approximately for $\Omega > \pi/4$).  For
similar background curvatures, but much larger ellipticities, the
potential develops a repulsive barrier before becoming attractive
for certain $\Omega$. This signals that $A_{4} < 0$ for large
enough $\D$ and is depicted in Fig. \ref{UOm}b for $\D = 2.5$. In
the limit of small $K_{b}$,  $A_{4} < 0$ when 

\begin{equation}
\D > \D^{*} + {K_{b}^{2}\over 8}\left(3+{\sqrt{2}\over
2}(3+\sin^{2}2\Omega)\right) + O\left(K_{b}^{4}\right).
\end{equation}

\noindent There is yet an additional, qualitatively different feature
of $U_{\em eff}(R;\D,K_{b},\Omega)$ when both  $\D$ and $K_{b}$ are
large.  Although typical values of $K_{b}$ (see Eq.  \ref{RESCALING})
in biological settings is $K_{b} \ll 1$, we find that large values of
$K_{b}$ and $\D$ give rise to {\it double} minima in the interaction
potential, especially near $\Omega \simeq \pi/2$.  Figure \ref{UOm}c
shows double minima for $\Omega = 7\pi/16, \pi/2$. Additional higher
order coefficients such as $A_{6}/R^{6}$, {\it etc.} are required to
quantitatively describe multiple minima. The two minima are a
consequence of the two independent physical effects that prefer energy
minima; local Gaussian curvature associated with $K_{b}$ and effective
ellipticity $\D$.  Typically, the weaker, longer-ranged minimum is
predominantly the signature of a large $K_{b}$, while the deeper,
shorter-ranged minimum (such as that shown in Fig. \ref{RMIN0/UMIN0}a and
\ref{UOm}b) is a feature of ellipticity $\D>\D^{*}$.  Saddles of order
$K_{b} > 1$ correspond to principle radii of curvature on the order of
$\sim 10$ times the protein size $a$, and are thus regions of extreme
Gaussian deformations. Regions of such warp may would only exist as
transient, small systems such as fusion necks. Henceforth, we will
restrict ourselves to $K_{b}$ small enough to only induce one minimum.

Angles $\Omega$ which yield attractive interactions can be estimated by
considering $A_{2}, A_{4}$. Assuming $A_{4} > 0$, values of $A_{2}<0$
give attractive interactions when $-\pi/4 < \Omega < \pi/4$.  When
$A_{2}>0$, proteins with orientation $\pi/4 < \vert\Omega\vert < 3\pi/4$
will experience attractive forces. However, these conditions are modified
if $A_{4}<0$, when some angles within $-\pi/4 < \Omega < \pi/4$ can yield
attraction even if $A_{2}>0$. This case corresponds to Fig. \ref{UOm}b
where a repulsive barrier at $R>R^{*}$ arises.  A  minimum can still
arise even at angles where $A_{2}\cos 2\Omega > 0$ due to the $-R^{-4}$
behavior.  The matching to repulsive behavior at smaller $R$ requires
consideration of $+R^{-6}$ terms.

\begin{figure}
\begin{center}
\vspace{-15mm}
\leavevmode
\epsfxsize=4.4in
\epsfbox{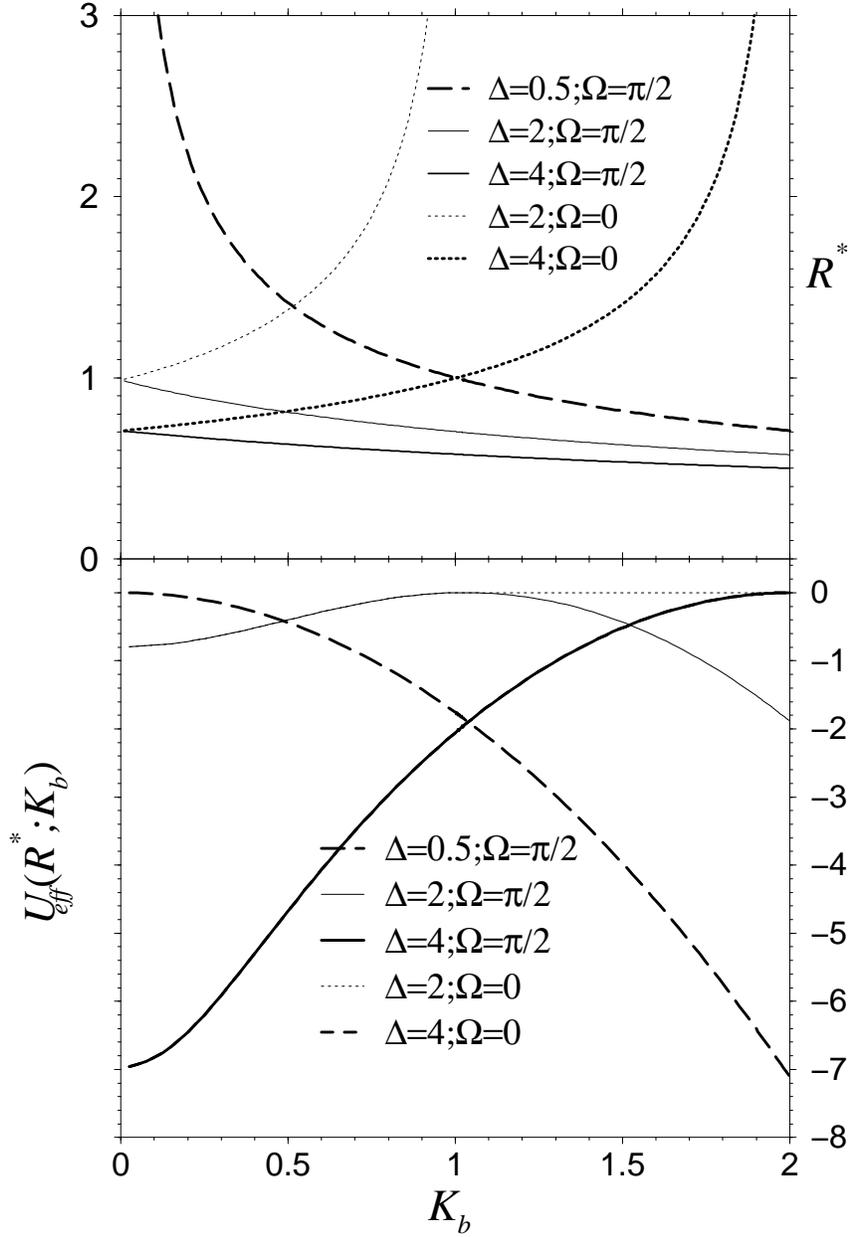}
\end{center}
\caption{(a). The radii corresponding to interaction potential minima as a
function of $K_{b}$ for $\D = 0.5,2,4$ and  $\Omega=0,\pi/2$. 
Curves that diverge signal a loss of the minimum 
(minimum radius $R^{*}\rightarrow \infty$) 
for parameters beyond those indicated. (b). The corresponding potential
energy well depths at $R^{*}$. The energies asssociated with 
$\D=2; \Omega=0$ and $\D=2; \Omega=\pi/2$ separate at 
$K_{b}\approx 1.1$ when the $\Omega=0$ energy well disappears. The minimum energies 
associated with large $\D$ and $\Omega=\pi/2$ is still 
increasing for  $K_{b}\gtrsim 1.1$.}
\label{ruKb}
\end{figure}

The top panel of Figure \ref{ruKb} shows the radius corresponding to the
only minimum of the effective potential $U_{\em eff}$ as a
function of $K_{b}$, for $\D = 0.5,2$ and $4$. Both east-west and
north-south configurations are shown, with intermediate angles $\Omega$
interpolating between the curves. For small ellipticity, the local principle curvature
$K_{b}$ is the predominant source of attraction at larger distances,
shown by the thick dashed curve. Increasing $K_{b}$ destabilizes
the effective energy minima near $\Omega = 0$. Above a certain background
Gaussian curvature intensity, the effective potential minimum evaporates
to $R^{*}\rightarrow \infty$ for proteins situated at $\Omega = 0$ (solid
curves), and the attraction is washed out. For small $K_{b}$, the two
effects, ellipticity and background Gaussian curvature, complement each
other near $\Omega = \pi/2$ in reinforcing an energy minimum.  Consistent
with Fig. \ref{RMIN0/UMIN0}a for $\D > \sqrt{2}$, $R^{*}$ in Fig.  \ref{ruKb}
(thick curves) is smaller for larger $\D$.  The bottom panel plots the
corresponding minimum energies.

\begin{figure}
\begin{center}
\leavevmode
\epsfxsize=4.4in
\epsfbox{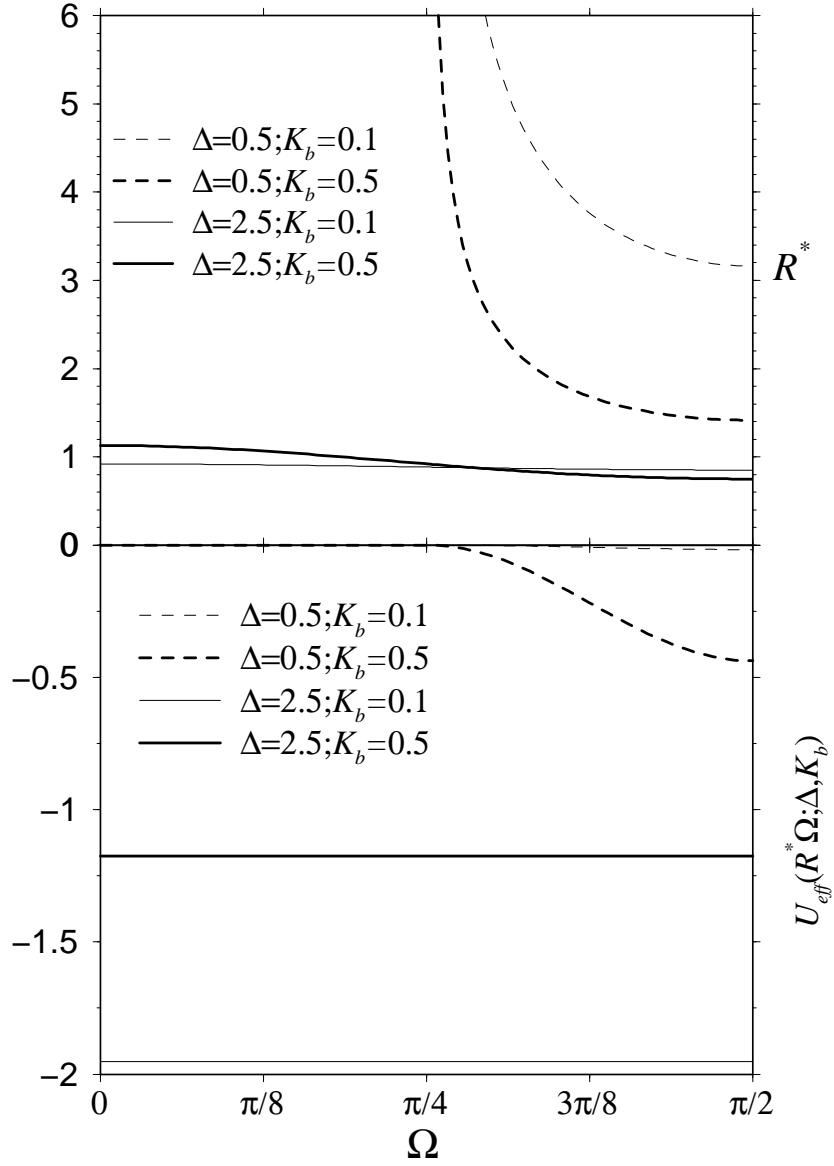}
\end{center}
\caption{Angular dependence of (a). $R$, and (b). $U_{\em eff}(R^{*},\Omega)$
as functions of pair orientation angle $\Omega$. Minima arising mainly from 
background saddle (sensitive to $\Omega$) and ellipticity (insensitive to $\Omega$) 
are shown.}
\label{ruOm}
\end{figure}

The $\Omega$-dependence of $R^{*}$ and the minimum energy is shown in
Figure \ref{ruOm}. As expected, or large $\D \gg \sqrt{2}$, both $R^{*}$
and $U_{\em eff}(R^{*},\Omega)$ are fairly insensitive to $\Omega$. When
$\D$ is small, the energy minima and their associated radii $R^{*}$,
caused predominantly by $K_{b}$, are very sensitive to orientation
$\Omega$. These behaviors are consistent with the energy profiles shown in
Fig. \ref{UOm}b.  In fact, for small enough $\D$, the minima near $\Omega
\approx 0$ are annihilated, independent of $K_{b}$.  Thus, we see a
qualitative difference between attractive potentials generated by
intrinsic ellipticity and background Gaussian curvature.

\section{The second virial coefficient}


We now consider the influence of the effective protein-protein
attractions on a low density ensemble of inclusions. By analogy with the
molecular origins of the osmotic second virial coefficients of proteins
in solution \cite{NEAL}, we will consider the bending energy
contributions to the second virial coefficient for a two-dimensional
protein equation of state. The membrane mediated interactions however,
are much longer-ranged than those in solution \cite{NEAL}.
Consider the thermodynamic limit and times long enough such that

\begin{equation}
\tau \gg {\ell^{2}\over D_{\em trans}} \gg \tau_{\em rot}, 
\end{equation}

\noindent where $D_{\em trans}$ is the protein translational diffusion
constant.  On the time scale $\tau$, the inclusions are relatively free to
diffuse about the bilayer. They interact among themselves via the
rotationally averaged potential $U_{\em eff}$ that manifests itself on
time scales $\gtrsim\tau_{\em rot}$.  For very low protein densities
(large protein separation $\ell$), the two dimensional protein osmotic
pressure will be nearly that of an ideal gas, analogous to a low density
gaseous phase surfactant monolayer at the air water interface.  Finite
protein size $a$, and longer-ranged elastically-coupled interactions will
give nonideal gas properties. The first correction to ideality in the
equation of state is given by the second virial coefficient 
\cite{MCQUARRIE}:

\begin{equation}
{\Pi \over k_{B}T} = \Gamma + B_{2}\Gamma^{2}  
+ O\left(B_{3}\Gamma^{3}\right), 
\label{EQS}
\end{equation}

\noindent where $\Gamma$ is the surface concentration of 
protein and $B_{2}$ is computed using the formula

\begin{equation}
\begin{array}{rl}
\displaystyle  B_{2}(\D,K_{b})&\equiv \displaystyle  -{1\over
2Z}\int_{0}^{2\pi}\int_{0}^{\infty}\left(e^{-E(R,\theta_{1},\theta_{2};\D,K_{b},\Omega)+
E_{\em eff}(\infty)}-1\right)RdR d\Omega d\theta_{1} d\theta_{2} \\[13pt]
\:\displaystyle &=\displaystyle -{1\over 2}\int_{0}^{2\pi}\int_{0}^{\infty}
\left(e^{-U_{\em eff}(R;\D,K_{b},\Omega)+U_{\infty}(\D,K_{b})}
-1\right)RdR d\Omega, 
\end{array}
\label{B2}
\end{equation}

\noindent The second virial $B_{2}$ represents the small fraction of
pairwise interacting proteins. Equations \ref{EQS} and \ref{B2} are
nondimensionalized such that the surface density $\Gamma \ll 1$ is
measured by the number of proteins in area $R_{0}^{2}$ (see Eq. 
\ref{RESCALING}) and the protein osmotic pressure $\Pi$ is measured in
units of $k_{B}T/R_{0}^{2}$. Equation \ref{B2} is {\it exact} and does
not require the separation of rotational and translational diffusion
times needed for the derivation of $U_{\em eff}(R;\D,K_{b},\Omega)$. 
Here, we do not consider how integrating out the rotational degrees of
freedom affect the fixed translational degree of freedom. Instead, we
are considering times long enough for equilibration of both degrees of
freedom, and their combined contribution to the equation of state via
$B_{2}$.

The physical origin and value of $K_{b}$ used in Eq.  \ref{B2} is as
follows.  The local curvature felt by the interacting pair represents
an interaction between this pair and some other distant proteins. 
However, the virial equation of state (Eq.  \ref{EQS}) is a systematic
expansion in surface density expanded about an ideal, {\it
noninteracting} ensemble.  Since membrane bending mediated
interactions are not pairwise additive \cite{KIM}, one might be
tempted to assume that the presence of other proteins would modify the
interaction energy $E$ used in the expression for $B_{2}$.  However,
these more complicated interactions would depend upon the
concentration of the other background proteins, and would generate
terms higher order in $\Gamma$.  In other words, we start at
densities so low that the protein ensemble is completely noninteracting.
As the density is slightly increased, a pair of protein molecules
occasionally interact and perhaps form dimers, with each pair ignorant
of any other protein. At this still rather low density, the
probability three or more proteins approach each other is negligible. 
When the density is further increased, one needs to consider the
higher order virial terms.  Therefore, to second order in $\Gamma$,
the deviation of the equation of state from ideality is completely
determined by the two-body interaction
$E(R,\theta_{1},\theta_{2};\D,\bar{K}_{b},\Omega)$ and is independent
of nonpairwise effects \cite{MCQUARRIE}.  Note however, that the
two-body interaction {\it will} depend on the $\bar{K}_{b}$ associated
with {\it externally} forced, zero mean curvature membrane
deformations.  Therefore, for the expansion Eq. \ref{EQS} to be
consistent, the value of $K_{b} = \bar{K}_{b}$ to be used in Eq.
\ref{B2} is that owing solely to external force-generated Gaussian
curvatures, independent of the protein density.

\begin{figure}
\begin{center}
\leavevmode
\epsfxsize=4.8in
\epsfbox{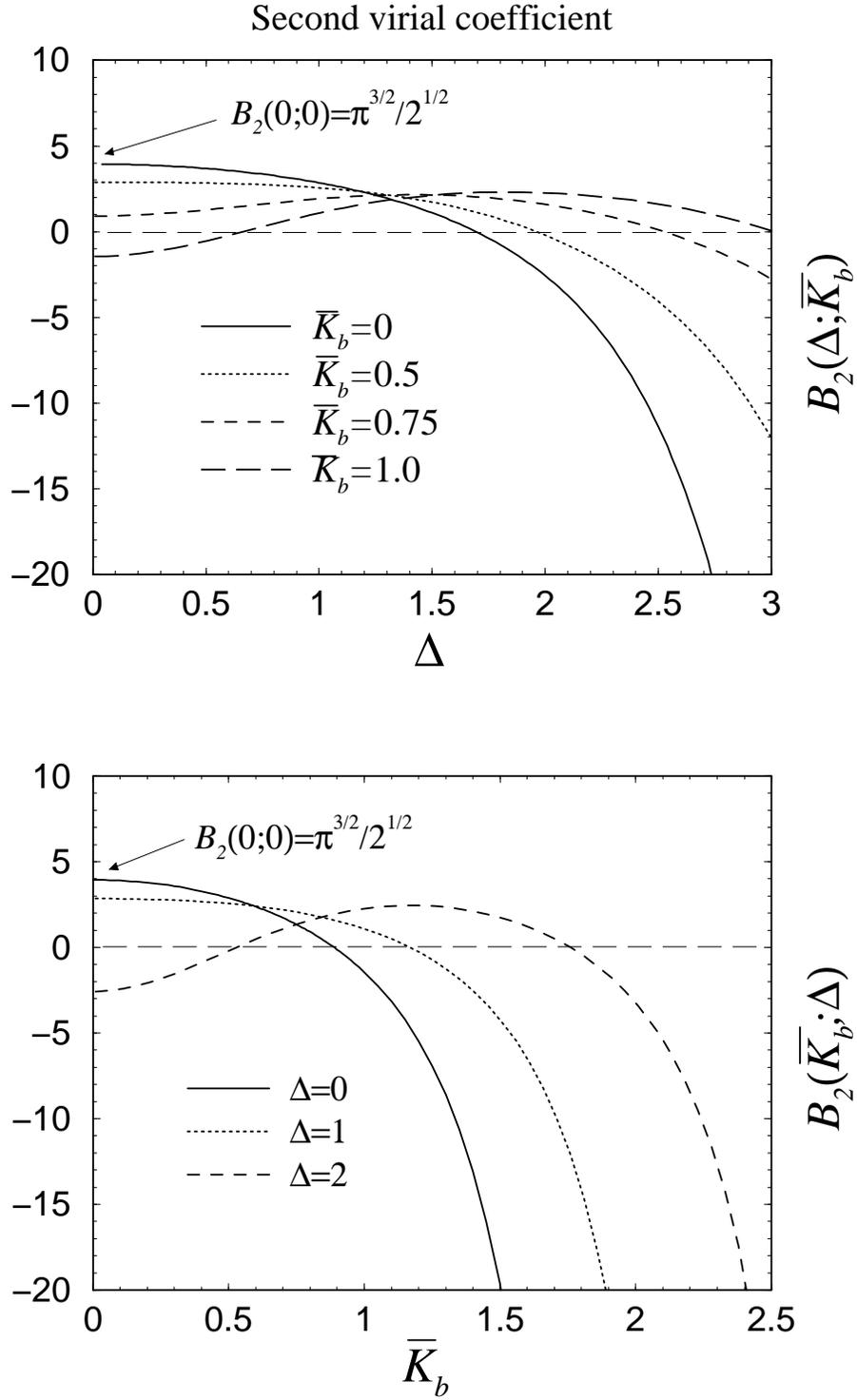}
\end{center}
\caption{(a). Second virial coefficient $B_{2}(\D,K_{b}=0,0.5, 0.75, 1)$. 
A negative virial
coefficient is indicative of an overall 
attractive interaction such that 
the osmotic pressure is {\it reduced} 
from that expected in ideal solutions. 
The value $B_{2}(0,0) = \pi^{3/2}/\sqrt{2}>0$ corresponds to 
the virial coefficient of circular, repulsive ($U=2/R^{4}$)
inclusions. (b). $B_{2}$ as a function of background 
saddle for various ellipticity parameters $\Delta$.}
\label{B2Kb=0}
\end{figure}

Figure \ref{B2Kb=0}a shows the numerically computed second virial
coefficient as a function of inclusion ellipticity for various
$\bar{K}_{b}$. As expected for small $\bar{K}_{b}$, the virial
coefficient becomes increasingly negative as the ellipticity
increases.  The value for circular inclusions
$B_{2}(\D=0,\bar{K}_{b}=0) = \pi^{3/2}/\sqrt{2}$ corresponds to purely
repulsive disks with mutual interaction $U(R) = 2/R^{4}$. At
ellipticity $\D \simeq 1.69$, $B_{2}(1.69, \bar{K}_{b}=0) \simeq 0$
corresponding to a protein solution that is ideal to second order in
surface density. Although when $\D\simeq 1.69 > \D^{*}=\sqrt{2}$,
$U_{\em eff}$ has an attractive minimum, its effects are negated by
the repulsive $R^{-4}$ part of the interaction such that the overall,
effective contribution to $B_{2}$ vanishes.  For $\D > 1.69$, the
effective attraction between membrane proteins begins to manifest
itself. The second virial is modified by externally imposed Gaussian
curvature. Recall that when $\bar{K}_{b} \neq 0$, certain angles
$\Omega$ lead to attractive interactions, even for small $\D <
\D^{*}$.  Since we are now thermodynamically averaging over protein
positions and $\Omega$ in addition to $\theta_{1},\theta_{2}$, the
inclusions will spend more time at attractive, lower energy angles
$\Omega$, hence lowering $B_{2}$.  Consistent with Fig. \ref{UOm},
larger values of $\bar{K}_{b}$ for $\D > \D^{*}$ lead to stronger
repulsions at small $\Omega$, which average into $B_{2}$, making it
less negative.

The dependence of $B_{2}$ on $\bar{K}_{b}$ is indicated in 
Figure \ref{B2Kb=0}b. In the absence of ellipticity,
$B_{2}$ is given by the integral

\begin{equation}
B_{2}(\D=0;\bar{K}_{b}) = -
\pi\int_{0}^{\infty}\left[e^{-2/R^{4}}I_{0}(4\bar{K}_{b}/R^{2})-1\right]
R dR.
\end{equation}

\noindent For $\D > 0$, $B_{2}$, found from numerical integration 
of the full expression Eq. \ref{B2}, are also shown in Fig. \ref{B2Kb=0}b.
For $\bar{K}_{b}=0$, increasing ellipticity decreases 
inclusion repulsions and $B_{2}$. As in Fig. \ref{B2Kb=0}a, large 
$\bar{K}_{b}$ and $\D$ tend to increase $B_{2}$.

Equation \ref{U2} was used in Eq. \ref{B2} to compute the curves shown in
Figures \ref{B2Kb=0}; thus, the protein-protein interaction was assumed
to consist of contributions only from membrane bending.  The hard core,
excluded area of each protein, $\sim \pi a^{2}$, can be included by
modifying $U_{\em eff}(R)$ by setting $U_{\em eff}(R\leq a/R_{0}) =
\infty$. Although we expect this additional repulsive term to further
reduce the effective sampling area of the inclusions, and increase the
second virial coefficient, we find that for all reasonable values of
$R_{0}$, $B_{2}$ does not change noticeably from those shown in Figs.
\ref{B2Kb=0}.  The hard core part of the potential, due to {\it e.g.}
close-ranged van der Waals repulsion, is not statistically sampled by the
inclusions since the membrane bending induced interactions ($\sim
1/R^{4}$) already keeps them far apart.

Since nonpairwise interactions manifest themselves 
only at third and higher order in $\Gamma$, we can estimate their
importance by comparing $B_{2}\Gamma^{2}$ with 
$B_{3}\Gamma^{3}$. For nonpairwise interactions to be 
thermodynamically relevant it is necessary but 
not sufficient that the surface density

\begin{equation}
\Gamma \gtrsim \Big|{B_{2}\over B_{3}}\Big|.
\end{equation}


\noindent Although multibody interactions may be important
microscopically, their effects on the low density equation of state,
cannot be resolved.  Even if the density is high enough for
$B_{3}\Gamma^{3}$ to be measurable, the value of $B_{3}$ is found via a
four-dimensional integral over configurations of three membrane proteins.
All orientations and distances will be averaged and all components of
their interactions, repulsive, attractive, pairwise, and nonpairwise will
be included. In other words, one cannot uniquely determine the potential
$U$ from a measurement of $B_{n}$.  

\section{Discussion and Conclusions}

Proteins beyond the range of screened electrostatic, or van der Waals
molecular forces can exert forces on one another by virtue of the
deformation they impose on the lipid bilayer.  These interactions can
be attractive if the proteins have a noncircular cross sectional shape
or if the local membrane deformation is saddle shaped (negative
Gaussian curvature).  For bending rigidities $b\approx 30k_{B}T$, and
protein shape ellipticities $\e/a \sim 0.25$, we find attractive
interactions of a few $k_{B}T$ acting at a range of $\sim 5$ protein
diameters.  Thus proteins of radii $\sim 5$nm can interact at
distances of $\sim 50$ nm, much further than any interaction between
similar molecules in solution. On a flat membrane ($H_{b}=K_{b} = 0$),
an effective ellipticity $\bar{\e}  > (2 k_{B}T/\pi b)^{1/2}$ is
necessary for a potential minimum to emerge between a pair of
proteins.  

Although we have presented the model in terms of integral membrane
proteins,  noncircular peripheral proteins, or
lipid/cholesterol/proteins aggregates can also induce local membrane
bending. Elastically-coupled interactions between peripheral membrane
proteins can mediate dissociation and binding. Dimerization of
noncircular peripheral proteins lowers their absolute energy below that
of separated ones, and so they are less likely to dissociate from the
membrane. Similarly, in one dimension, adsorbed proteins can bend DNA and
affect the binding of a second nearby protein. The effective interactions
in that case also depends upon the protein orientation
\cite{DIAMANT,DNA}.  Moreover, proteins and protein aggregates need not
be rigid, as we have assumed here.  Noncircular distortions of a lipid
or protein domains can fluctuate in such a way as to yield
domain-domain interactions of exactly the same form considered here.

Elastically mediated attractions can also manifest themselves in the
aggregation of {\it circular} proteins.  Once circular proteins overcome
short-ranged repulsions and dimerize, barriers to further aggregation of
these elliptical dimers are reduced by dimer-dimer attractions described
by Eq. (\ref{UEFF}). If the inclusions are themselves dimers or higher
aggregates that persist on the time scale of rotation, bending mediated
attraction would enhance further aggregation.  

We have only considered the mechanical energies of the intervening
lipid bilayer.  However, an ensemble of membrane-bound proteins or a
mixture of lipids can manifest itself through other forces. For
example, the presence of charged membrane components can induce
bending \cite{CHOU} and initiate endo/exocytosis and organelle
trafficking. Protein-protein interactions  arising from screened
electrostatic forces operate at much shorter distances than those of
the bending induced forces. Therefore, by spatially organizing charged
membrane components, elastic interactions may also play an indirect
role in large scale electrostatically induced membrane deformation.

We also considered an ensemble of surface proteins elastically coupled
by membrane deformation and computed the deviation of its equation of
state from that of an ideal solute.  Although membrane-mediated
protein-protein interactions are nonpairwise additive \cite{KIM}, only
the two-particle interaction is relevant for sparsely distributed
proteins.  On a flat membrane the second virial coefficient $B_{2} <
0$ when $\bar{\e} \gtrsim 0.95\sqrt{k_{B}T/b}$ ($\D \approx 1.69$). At
this ellipticity, the elastically induced $1/r^{4}$ repulsive
interactions just compensate for the rotationally averaged
attractions.  This dependence on $b/k_{B}T$ suggests that the cell can
regulate protein-protein interactions by varying the lipid composition
and hence the bending rigidity of the bilayer. Thus, the formation of
cholesterol rafts and lipid domains may have an indirect role in
mediating long-ranged surface protein aggregation and activity.  

Finally, we propose possible experiments in artificial membrane
systems where the surface density can be made small enough for a
virial expansion to be valid. Although the two-dimensional osmotic
pressure would be difficult to measure accurately, measurements of the
association time between dimerized proteins are feasible. 
Measurements have been made of the lifetimes of gramicidin A channels
composed of dimers of barrels in opposite bilayer leaflets as a
function of bilayer thickness \cite{gA1,gA2}.  Measurements of dimer
lifetimes as a function of lipid tail length may reveal the dependence
of the attractive interactions outlined in this paper. In fact, since
the second virial coefficient measures the time-averaged fraction of
proteins in dimers at low density, their lifetimes are proportional to
$B_2$ for attracting proteins. Moreover, these association lifetimes
can be measured in the presence of externally imposed Gaussian
deformations.  Even though an imposed Gaussian curvature increases the
interaction well depth at $\Omega \approx \pi/2$, and destroys the
attractions for proteins near $\Omega \approx 0$, the overall
statistical effect, is to enhance binding, as is evident from Figures
\ref{B2Kb=0}.  Therefore, we expect that dimer lifetimes can be
enhanced for proteins residing in regions of large magnitudes of
Gaussian curvature such as the base of extruded tubules.  This may be
instrumental in recruiting fusagens to the correct location for
membrane budding. Dimer lifetimes potentially can be measured by
fluorescence transfer of specifically designed hydrophilic moieties
attached to membrane proteins.  Nonpairwise interactions can only be
probed directly by measuring lifetimes and aggregation rates of
trimers.  This would require statistical analyses of chemical or
fluorescence activity among two differently tagged membrane proteins,
or single molecule diffusion studies.

\vspace{5mm}

We thank J. B. Keller and J. C. Neu for helpful comments and many enlightening
discussions. T. C.  acknowledges support from the National Science Foundation via grant
DMS-9804370.  K. K. is supported by a grant from the Wellcome Trust, and  G. O. is
supported by NSF grant DMS-9220719.

\begin{appendix}

\section{Interaction energy among noncircular inclusions}

We consider the boundary conditions that the height, $h(r,\theta)$,
must satisfy and the effects of noncircular proteins on the
interaction energies \cite{KIM2}.  Consider proteins with chemistry
that changes the cross-sectional protein shape from circularity by an
amount $\e$.  The concomitant changes in lipid contact height and
angle are also assumed to be modified by $O(\e)$. As shown in Fig.
\ref{FIG1}, the protein perimeter, measured from the protein center is,
to order $O(\e)$, 

\begin{equation}
\C \simeq (a+\e\cos 2(\t-\t_{i})){\bf n},
\end{equation}

\noindent where ${\bf n}$ is the unit normal vector to the curve ${\bf
C}$ projected onto the bilayer midplane, and $\e\cos 2(\t-\t_{i})$ is a
small, angle-dependent perturbation measuring the deviation from
circularity of protein $i$. Upon expanding the general boundary
conditions $h(\C) = \d h(\t)$ and ${\bf n}\cdot\nabla h(\C) = -\gamma
- \d\g(\t)$ to lowest order in $\e$, we arrive at effective boundary
conditions:

\begin{equation}
\begin{array}{l}
h(a) \simeq  \d h(\t-\t_{i})+\g\e\cos 2(\t-\t_{i})+O(\e^{2})\\[13pt]
\displaystyle \partial_{r} h(a) \simeq -\g\left(1+{\e\over a}\cos
2(\t-\t_{i})\right)-\d\g(\t-\t_{i}) 
+O(\e^{2})
\end{array}
\label{BC}
\end{equation}

\noindent where we have for simplicity also assumed the
variations $\d h(\C)$ and $\d \g(\C)$ to be also of order $\e$.

In the limit of small noncircularity or low protein concentrations,
the dominant nondivergent contribution of $H({\bf r})$ to the energy
$\tilde{E}$ is $a_{2}^{2}+b_{2}^{2}$. The deformation $h(r,\theta)$
that satisfies $\nabla^{2}h(r,\theta) = 2H(r,\theta)$ and Eqs.
\ref{BC} can be written in the form

\begin{equation}
h(r,\theta) \simeq -
\gamma\ln\left({r\over a}\right) + \sum_{n=2}^{\infty}\left(f_{n}(r)\cos n\theta
+g_{n}(r)\sin n\theta\right),
\label{MULTIPOLE}
\end{equation}

\noindent and determine $a_{2}, b_{2}$. When the proteins
have intrinsic noncircularity ($\varepsilon \neq 0$),
$a_{2}^{2}+b_{2}^{2}$ turns out to be the magnitude of the
local Gaussian curvature (since $H_{b} = 0$), modified by additional
$\theta_{i}-$dependent terms \cite{KIM2}.  The local Gaussian
curvature due to the other $j$ far field proteins, in either case, is
calculated using the leading order term  $h(\vec{r}) \approx
-\gamma \ln \vert \vec{r}-\vec{r}_{j}\vert$, which is simply a
superposition of the longest-ranged $\ln r$ terms about each
inclusion. The total bending energy
$\tilde{E}[H(r,\theta)]$ for an ensemble of $N$ inclusions
can be written in the complex form \cite{KIM2},

\begin{equation}
\tilde{E} = \pi b \gamma^{2}\sum_{j}\Big|\sum_{i\neq j} {a^{2}
\over (z_{i}-z_{j})^{2}}-{\bar{\e}\over
2}e^{-2i\theta_{j}}\Big|^{2}.
\label{Eetab}
\end{equation}

\noindent where $z_{i}=x_{i}+iy_{i}$ is the position of the
$i^{th}$ protein in the complex plane, and

\begin{equation}
\bar{\e}\equiv \left({\e \over a}\right)
\left(\gamma+2{\d h \over a} -\d
\gamma \right)
\end{equation}

\noindent measures the effective ellipticity of the identical proteins. 
Now consider two relatively isolated, identical proteins $i,j=1,2$. The
effects of proteins far away are felt via a local Gaussian curvature
emanating from these background proteins. Upon explicitly separating 
these contributions, the pair interaction energy becomes

\begin{equation}
\tilde{E}(r,\theta_{1},\theta_{2};\eta_{b}, \Omega) 
= \pi b \left[\Big| {a^2 \gamma e^{-2i\Omega} \over r^{2}}+\eta_{b} 
- {\bar{\e}\over
2}e^{-2i\theta_{1}}\Big|^{2} + \Big|{a^2 \gamma e^{-2i\Omega}
\over r^{2}}+\eta_{b} - {\bar{\e}\over
2}e^{-2i\theta_{2}}\Big|^{2}\right]
\label{U1}
\end{equation}

\noindent where

\begin{equation}
\eta_{b} \equiv a{\partial^{2} h_{b}(\S)\over \partial x_{1}^{2}}
= -a{\partial^{2} h_{b}(\S)\over \partial x_{2}^{2}}
\end{equation}

\noindent is the curvature in the ${\bf x}_{1}$ principle direction
due to far-field background inclusions or externally induced
deformations $h_{b} \approx -\gamma \ln \vert z-z_{j}\vert,\, j\geq
3$. The mean curvature expanded about a noncircular protein (Eq. 
\ref{KAPPAMP}) results in a deformation $h(r,\theta)$ with terms
proportional to $r^{2}\cos2\theta, r^{2}\sin 2\theta$ \cite{KIM}. 
These term carry zero mean curvature, but constant negative Gaussian
curvature. From the expansion Eq.  \ref{KAPPAMP}, the only mean
curvature contribution can be seen to decay as $r^{-2}$, which we
neglect. A further contribution to the local saddle curvature,
$\eta_{b}^{2}$, felt by the two proteins can arise from externally
applied mechanical forces that deform the bilayer in an appropriate
way.  The angles $\theta_{1},\theta_{2}$ are the angles of the
principle axes of the inclusion shape (or the height or contact angle
functions $\d h, \d \gamma$) measured from the principle background
curvature axis ${\bf x}_{1}$. The angle $\Omega$ measures the angle
between the principle background curvature axis and the segment
joining the centers of the two inclusions. Upon rescaling according to
Eq.  \ref{RESCALING}, we arrive at the energy given in Equation
\ref{U2}.

\section{Rotational averaging}

The integrals 

\begin{equation}
\int_{0}^{2\pi} E(R,\t_{1},\t_{2};K_{b},\Omega)e^{-E} d\t_{1} d\t_{2}
\quad \mbox{and}\quad Z^{1/2} \equiv \int_{0}^{2\pi} e^{-E} d\t_{1} d\t_{2}
\end{equation}

\noindent used to compute the rotationally averaged, effective 
protein-protein interaction involve integration of

\begin{equation}
\int e^{\a\cos 2\t + \b\sin 2\t} d\t\,\,\,\,\mbox{and}\,\,\,
\int (\a\cos 2\t + \b\sin 2\t)e^{\a\cos 2\t + \b\sin 2\t} d\t.
\label{ab}
\end{equation}

\noindent The first integral in Eq. \ref{ab} can be computed 
in closed form by substituting the exponents with their 
Bessel function expansions

\begin{equation}
\begin{array}{l}
\displaystyle e^{\a\cos 2\t} = I_{0}(\a) + 
2\sum_{n=1}^{\infty}i^{n}I_{n}(\a)\cos 2n\t \\[13pt]
\displaystyle e^{\b\sin 2\t} = I_{0}(\b) + 2\sum_{n=1}^{\infty}(-1)^{n}I_{2n}(\b)\cos
4n\t - 2\sum_{n=1}^{\infty}i^{2n+1}I_{2n+1}(\b)\sin 2(2n+1)\t
\end{array}
\label{EXPANSION}
\end{equation}

\noindent and integrating term by term. The cross-terms of the product 
of the two equations in Eq. \ref{EXPANSION} involve single powers of 
$\cos$ and $\sin$ and vanish upon integration.  We are left with 

\begin{equation}
Z^{1/2} = 2\pi I_{0}(\a)I_{0}(\b) + 4\pi \sum_{n=1}^{\infty}
(-1)^{n}I_{2n}(\a)I_{2n}(\b).
\label{Z1/2}
\end{equation}

\noindent An analytic continuation of the sum formula,

\begin{equation}
J_{0}(\sqrt{\a^2+\b^2-2\a\b\cos\varphi}) = 
J_{0}(\a)J_{0}(\b) + 2\sum_{n=1}^{\infty}J_{n}(\a)J_{n}(\b)\cos n\varphi,
\end{equation}

\noindent at $\varphi = \pi/2$ simplifies Eq. \ref{Z1/2} to,

\begin{equation}
Z^{1/2} = 2\pi I_{0}(\xi), \quad \xi \equiv \sqrt{\a^2+\b^2}.
\end{equation}

Finally, the second integral in Eq. \ref{ab} can be computed
by taking derivatives of $Z^{1/2}$:

\begin{equation}
\int (\a\cos 2\t + \b\sin 2\t)e^{\a\cos 2\t + \b\sin 2\t} d\t =
\left(\a{\partial\over \partial\a} + \b{\partial\over \partial_{\b}}\right)Z^{1/2}.
\end{equation}

\noindent Using these results, we arrive at the rotationally
averaged energy $E_{\em eff}$ given by Eq. \ref{UEFF2}. For large 
separation distances $R$, the effective interaction $U_{\em eff}(R)\equiv 
E_{\em eff}(R)-E_{\em eff}(\infty)$ defined in Eq. \ref{UEFF} 
can be expanded as in Eq. \ref{EXPANSIONR} where the coefficients are given by

\begin{equation}
A_{2} \equiv 4 K_{b}-2\D^{2}K_{b}{\partial \over \partial \xi}\left(
{I_{1}(\xi)\over I_{0}(\xi)} \right)_{\D K_{b}}\!- 2\D {I_{1}(\D K_{b})\over 
I_{0}(\D K_{b})}
\label{A2}
\end{equation}

\noindent and 

\begin{equation}
\begin{array}{r}
\displaystyle A_{4} \equiv 2-\D^{2}{\partial\over \partial \xi}\left({I_{1}(\xi)
\over I_{0}(\xi)}\right)_{\D K_{b}}\!- {\D \over K_{b}}{I_{1}(\D
K_{b})\over I_{0}(\D K_{b})}\sin^{2}2\Omega - \D^{2}\left[K_{b}{\partial^{2}
\over \partial \xi^{2}} + {\partial \over \partial \xi}\right]\left(
{I_{1}(\xi)\over I_{0}(\xi)}\right)_{\D K_{b}}\!\cos^{2}2\Omega.
\end{array}
\label{A4}
\end{equation}

\end{appendix}

\newpage

\end{document}